\documentclass[fleqn,usenatbib]{mnras}

\usepackage{newtxtext,newtxmath}

\usepackage[T1]{fontenc}

\DeclareRobustCommand{\VAN}[3]{#2}
\let\VANthebibliography\thebibliography
\def\thebibliography{\DeclareRobustCommand{\VAN}[3]{##3}\VANthebibliography}

\usepackage{graphicx}	
\usepackage{amsmath}	
\usepackage{tabularx}

\title{Search for pulsars with periods of more than two seconds at declinations from $+21^{\circ}$ to $+42^{\circ}$}

\author[S. A. Tyul'bashev and G.E. Tyul'basheva]{
	S. A. Tyul'bashev $^{1}$\thanks{E-mail: serg@prao.ru}
	G.E. Tyul'basheva,$^{2}$
	\\
	$^{1}$ Pushchino Radio Astronomy Observatory, Astro Space Center, Lebedev Physical Institute, Russian Academy of Sciences, Pushchino, 142290 Russia \\
	$^{2}$ Institute of Mathematical Problems of Biology RAS (IMPB RAS) brunch of Keldysh Institute of Applied Mathematics of Russian Academy of Sciences, \\
	Pushchino, Moscow reg., Russia \\
}

\date{ }

\pubyear{ }

\begin{document}
	\label{firstpage}
	\pagerange{\pageref{firstpage}--\pageref{lastpage}}
	\maketitle
	
	\begin{abstract}
A search was carried out for pulsars with periods ($P$) from 2 to 90~s in daily observations carried out over an interval of 5 years in a area measuring 6300 sq.deg. The data was obtained on a Large Phased Array (LPA) at a frequency of 111 MHz. The periodograms calculated using the Fast Folding Algorithm (FFA) were used for the search. To increase the sensitivity, the periodograms obtained in different observation sessions were added together. Of the 14 known pulsars that entered the study area, with periods of $P>2$~s and dispersion measures ($DM$) less than 200 pc/cm$^3$, 9 were detected. 2 new pulsars have been found. The mean profiles of pulsars are obtained and estimates of their flux densities are given. The open pulsar J1951+28, with a period of $P = 7.3342$~s and $DM=3.5$~pc/cm$^3$, turned out to be one of the pulsars closest to the Sun. The absence of new pulsars with periods of tens of seconds with a search sensitivity of 1 mJy outside the Galactic plane indicates a low probability of the existence of pulsars with extremely long periods. Most likely, the recently found sources of periodic radiation with periods from a minute to tens of minutes are white dwarfs.

	\end{abstract}
	
	\maketitle 
	
	\section {Introduction}

As a rule, using the word radio pulsar (pulsar), it is meant that it is a rapidly rotating neutron star that emits a pulse in the radio range at each or almost every revolution. We see this pulse due to the fact that the axis of rotation and the magnetic axis are located at an angle to each other.

The first pulsars were discovered by observing individual pulses (\citeauthor{Hewish1968}, \citeyear{Hewish1968}), however, immediately after the discovery it became clear that in order to search for and study weak pulsars, it was necessary to increase the signal-to-noise ratio ($S/N$). To do this, individual pulses must be added together. Indeed, by adding pulses with a known period, we increase $S/N$ as the square root of the number of stacked pulses. Thus, by adding up the pulses over increasingly long time intervals, it is possible to detect fainter and fainter pulsars.

When searching for pulsars, two characteristics need to be selected - this is the pulsar period ($P$) and the dispersion measure ($DM$), reflecting the number of electrons on the line of sight. Simple addition of pulses with simultaneous iteration of all possible values of $DM$ turned out to be very expensive from the point of view of computational resources, therefore, almost immediately after the discovery of pulsars, fast summation algorithms, well-known in mathematics, were proposed to estimate $P$ and $DM$ when searching for pulsar candidates. The power spectra obtained using the Fast Fourier Transform (FFT) (\citeauthor{Lovelace1968}, \citeyear{Lovelace1968}) reveal periodic signals in the frequency domain. Periodograms obtained using the Fast Folding Algorithm (FFA) (\citeauthor{Staelin1969}, \citeyear{Staelin1969}) reveal periodic signals in the time domain. The use of power spectra to search for pulsars turned out to be low-cost in computing resources, and it is still used with small variations. Periodograms require more computational resources than power spectra, so their use for searching for pulsars began to be considered again almost 50 years after the first application (\citeauthor{Cameron2017}, \citeyear{Cameron2017}; \citeauthor{Morello2020}, \citeyear{Morello2020}).

Both power spectra and periodograms make it possible to search for pulsars, but there are nuances that need to be taken into account. Fourier spectra work better if the signal occupies a large part of the period, thereby becoming similar to a part of a sine wave. Periodograms work better with narrow signals. Unlike Fourier power spectra, in which the total pulse energy in a periodic signal is distributed over all harmonics, periodograms collect all the energy in one harmonic, which allows for maximum sensitivity.  Recent searches for pulsars using periodograms have shown that for pulsars with narrow pulses occupying a small part of the period, the sensitivity of the FFA search can be up to 7-8 times higher than the standard search using power spectra (\citeauthor{Parent2018}, \citeyear{Parent2018}; \citeauthor{Singh2023}, \citeyear{Singh2023}).

When working with real data rather than a modulated signal, the background signal must be subtracted from the observational data (baseline subtraction) before using FFT or FFA. The signal observed on the radio telescope consists of a confusion of extended and compact radio sources that fall into the radiation pattern of the radio telescope. The galactic background, frequency and time interference, and radiation from strong sources in the side and rear lobes of the antenna are superimposed on this confusion. In the meter wavelength range, scintillation is also added on the interplanetary medium and on the ionosphere. All components of the observed signal have different characteristic time scales, and baseline subtraction becomes a non-trivial procedure. As a result of incomplete subtraction of the baseline, components (red noise) corresponding to low frequencies (long periods) remain in the power spectra.

The search for slow pulsars on the Large Phased Array (LPA) radio telescope, conducted using summed power spectra, showed that high sensitivity can be guaranteed for pulsars with $P<2-3$~s (\citeauthor{Tyulbashev2024}, \citeyear{Tyulbashev2024}). At periods of $P>2-3$~s, low-frequency noise begins to appear, interfering with the visual search for pulsar candidates, and at periods of $P>3-4$~s, red noise does not allow harmonics to be distinguished. However, if the pulsar is strong, it can be detected by the second, third and subsequent harmonics.

The total number of pulsars with long periods is small, and their search using Fourier power spectra is difficult for the reasons mentioned above. Indeed, in the ATNF pulsar catalog\footnote{ https://www.atnf.csiro.au/research/pulsar/psrcat /} (\citeauthor{Manchester2005}, \citeyear{Manchester2005}) as of April 2024, about 3,000 radio pulsars are included, of which $P>2$~s have about 250 pulsars ($\simeq 8\%$), $P>4$~s is observed in 78 pulsars ($\simeq 2.5\%$), $P>6$~s - 21 pulsars ($\simeq 0.7\%$). Unlike power spectra, low-frequency noise in periodograms is less pronounced, so periodograms can be used to search for pulsars with long periods.

Over the past five years, there has been an unexpected increase in the list of pulsars with extra-long periods. The papers \citeauthor{Tan2018} (\citeyear{Tan2018}); \citeauthor{Caleb2022} (\citeyear{Caleb2022}); \citeauthor{HurleyWalker2022} (\citeyear{HurleyWalker2022}); \citeauthor{HurleyWalker2023} (\citeyear{HurleyWalker2023}) talk about the discovery of radio pulsars J0250+5854, J0901-4046, J1627-5235, J1839-1031 with periods of $P=23.5$~s, 75.88~s, 1091~s, 1318~s, detected at frequencies of 135 MHz (LOFAR), 1.3 GHz (MeerKAT), 150 and 80 MHz (MWA). However, it is not clear whether all of these pulsars are neutron stars, or some of them are white dwarfs. If the indicated sources with extra long periods are canonical pulsars, then their total number in the sample of radio pulsars should be extremely small, if they are white dwarfs, their number may be large (\citeauthor{Rea2024}, \citeyear{Rea2024}). To answer this question, observations of large celestial areas over long time intervals are needed.

In this paper, we conducted a search for pulsars with periods in the range of 2-90~s, realizing high sensitivity by summing FFA spectra, which allowed us to eliminate the low-frequency (red) noise observed in Fourier power spectra at periods of $P>2$~s.

\section{Observations and processing}

To search for pulsars, round-the-clock monitoring observations were used, conducted since 2014 on the radio telescope of the LPA Lebedev Institute of Physics (LPI) at a frequency of 111 MHz. The LPA LPI has two independent radiation patterns, one of which (LPA3) has 128 stationary beams that cover declinations of $-9^{\circ} < \delta <+55^{\circ}$ (\citeauthor{Shishov2016}, \citeyear{Shishov2016}; \citeauthor{Tyulbashev2016}, \citeyear{Tyulbashev2016}). Data is recorded in two time-frequency modes: 32 frequency channels 78 kHz wide and with a sampling of 12.5 ms, and 6 frequency channels 415 kHz wide and with a sampling of 100 ms. At low time-frequency resolution, the recording volume per year in 128 beams is slightly more than a terabyte, at high time-frequency revolution - about 42 TB. The effective area of the radio telescope is approximately 45,000 square meters, the reception band is 2.5 MHz. The beams of LPA3 intersect at a power level of 0.405.

The data is recorded on three recorders. We processed data from a recorder that recorded 48 LPA3 beams covering declinations of $+21^{\circ} < \delta < +42^{\circ}$. There are long-term series of observations on these declinations, and a minimum of interference is recorded on them. A five-year observation interval for 2015-2019 was taken for processing. At declinations below $+21^{\circ}$, the amount of industrial interference increases, and additional forces need to be taken when processing data. At declinations above $+42^{\circ}$, the recorders were connected to the LPA3 beams in 2021-2022, i.e. there is only a relatively short series of observations available.

The processing scheme repeats the previously implemented method of searching for pulsars using FFT (\citeauthor{Tyulbashev2022}, \citeyear{Tyulbashev2022}). The essence of the processing was that in order to increase sensitivity, we added up the power spectra corresponding to observations from one direction in the sky for the entire monitoring observation interval. This increased the sensitivity of the search by about 40 times when adding up the power spectra over an interval of about 3,000 days (observation period 2014-2022). The power spectra obtained over different days can be added together. If there is a pulsar in the recording, information about the pulse phase of the pulsar is lost in the power spectrum, but information about its period remains. And this period always corresponds to a certain point number on the Fourier power spectrum.

A similar situation occurs when summing periodograms. If there is a periodic signal from a pulsar in the recording, it is reflected in the FFA spectrum as a harmonic visible on the main period of the pulsar and periods of multiples of the main one. The phase of the signal is lost. Since the periods of pulsars change slowly over time, the harmonics corresponding to the periods of pulsars in the FFA spectra fall on the same numbers of points in these spectra. If there is a pulsar in some direction, when summing the periodograms for different days, the $S/N$ harmonics will grow.

The FFA programme described in \citeauthor{Morello2020} (\citeyear{Morello2020}) was used to process the observations\footnote{ https://github.com/v-morello/riptide}. As recommended by the authors of the program, the account was conducted independently for periods of $2 < P < 50$~s and $30 < P < 90$~s. The data was processed sequentially by day. Each day was divided by coordinates fixed by right ascension and declination. Each right ascension coordinate was processed in parallel on the server's multicore processor. The processing of a year of observations took 3 days.

To save computing resources, we took data with low time-frequency resolution. To a first approximation, the use of data with low time-frequency resolution is justified, since according to the ATNF catalog, pulse widths of pulsars at the level of 10\% of the profile width ($W_{10}$) with $P> 2$~s begin at 40 ms. For such pulsars, the loss of $S/N$ in the FFA spectrum due to too long sampling will be equal to $(100/40)^{1/2} = 1.6$. Half of the pulsars with $P > 2$~s have pulse widths of $W_{10} > 80$~ms, so the loss of $S/N$ for such pulsars will be less than $(100/80)^{1/2} = 1.1$ times. According to our estimates, the sensitivity when adding periodograms over a 1-year interval can increase by about 10-15 times (\citeauthor{Tyulbashev2020}, \citeyear{Tyulbashev2020}). This estimate of the $S/N$ growth is valid if the sampling of the point is equal to the pulse duration. Considering that we process data with low time-frequency resolution, the average sensitivity loss will be about 30-40\% and the sensitivity in the summed FFA spectra will increase by 7-11 times.

Since the pulse width may be longer than the sampling of the point, for such pulsars there will be a loss of sensitivity already due to too wide pulses. Therefore, FFA spectra must be obtained with different initial averaging of the data. We averaged the data in moving average increments of 2, that is, assuming a pulse width of 100 ms, 200 ms, 400 ms and up to 25600 ms. According to the ATNF, 90\% of all pulsars with a $W_{10}$ estimate have a ratio of $W_{10}/P$ that does not exceed 7.5\%. That is, for a pulsar with $P = 90$~s, the pulse width will be $W_{10} < 6.75$~s (6750~ms) for 90\% of pulsars. We are searching for significantly larger estimated pulse widths and therefore do not expect a decrease in sensitivity for pulsars with wide mean profiles. When viewing periodograms, the visualization program allows you to output $S/N$ in harmonics depending on the expected pulse width. This allows us to make a rough estimate of the pulse width.

Thus, using data with low time-frequency resolution can potentially lead to a loss of sensitivity of several tens of percent. We considered the possible loss of sensitivity acceptable, given that FFA data processing with high time-frequency resolution can take about a year of server operation, according to estimates. Our plans include searching for pulsars using periodograms based on data with high time-frequency resolution for pulsars with periods from 0.025~ms. However, the purpose of this work is to search for pulsars with $P > 2$~s and estimate their number in the sky.

The sensitivity of the telescope when searching for pulsars varies, and depends on the period and $DM$ of pulsar. According to Fig. 4 in \citeauthor{Tyulbashev2022} (\citeyear{Tyulbashev2022}), for pulsars with $P>2$~s, all pulsars with an integral flux density of $S_{int}> 10-13$~mJy and $S/N\ge 6$ will be detected in a 3.5-minute observation session. We have selected the search boundary of $S/N\ge 10$. Taking into account the expected increase in sensitivity over the annual interval, we obtain that the sensitivity limit during the search will be $S_{int} \simeq 1.1$~mJy in the direction of the zenith and outside the plane of the Galaxy. Since the observations were not carried out at the zenith, and the sensitivity between the LPA beams may drop by $0.405^{-1}$=2.5 times (\citeauthor{Shishov2016}, \citeyear{Shishov2016}), the worst sensitivity will be 2.7 mJy, taking into account the cosine of the source height above the horizon at declination $+21^{\circ}$ and hitting the pulsar exactly between the beams. Thus, 2.7 mJy is the expected limit of the completeness of the survey. We need to register all pulsars that have $S_{int} \ge 2.7$~mJy. At the same time, the weakest pulsars available for detection may have $S_{int} = 1.1$~mJy. For the Galactic plane, sensitivity estimates will deteriorate by about 2 times. All sensitivity estimates presented relate to $DM <200$~pc/cm$^3$ (\citeauthor{Tyulbashev2022}, \citeyear{Tyulbashev2022}). At high values of $DM$, the sensitivity drops sharply due to scattering.

The total periodograms calculated for a given direction in the sky with an enumeration of various $DM$s are presented on maps of the form $P/DM$. Despite the fact that no pulsar with $DM > 250$~pc/cm$^3$ (\citeauthor{Tyulbashev2024}, \citeyear{Tyulbashev2024}) was found in the PUMPS pulsar search project, maps were built up to $DM = 1000$~pc/cm$^3$. Such a high $DM$ limit is due to the fact that the longer the pulsar period, the wider the pulses are on average, and the less the pulse scattering affects the sensitivity. Therefore, the maps are built with a margin of $DM$, in case pulsars with wide profiles and long periods are detected.

To obtain $P/DM$ maps in each FFA spectrum, harmonic values having $S/N$ greater than the specified value are stored, and only these values are shown on the maps. The pulsar on the $P/DM$ map is visible as a set of vertical segments bounded by $DM$ (y-axis) and having coordinates along the $P$ (x-axis) close to the pulsar period and to multiple periods (see Fig.1). The values of the $S/N$ periodograms are presented on an interactive the $P/DM$ map in the visualization program. Using the mouse, you can capture the values of $P$ and $DM$ on the segments and see the original periodograms, as well as more accurately estimate $DM$. Characteristic triangular structures are visible on the summed FFA spectrum (middle panel), the maxima of which coincide with the fundamental period and its multiples of harmonics. The right panel allows you to roughly estimate the expected $DM$ of the pulsar. Since the pulsar is very strong, its signal in the periodogram is visible on all $DM$s. The horizontal blue line in the figure marks the level of $S/N = 6$.

\begin{figure*}
	\includegraphics[width=\textwidth]{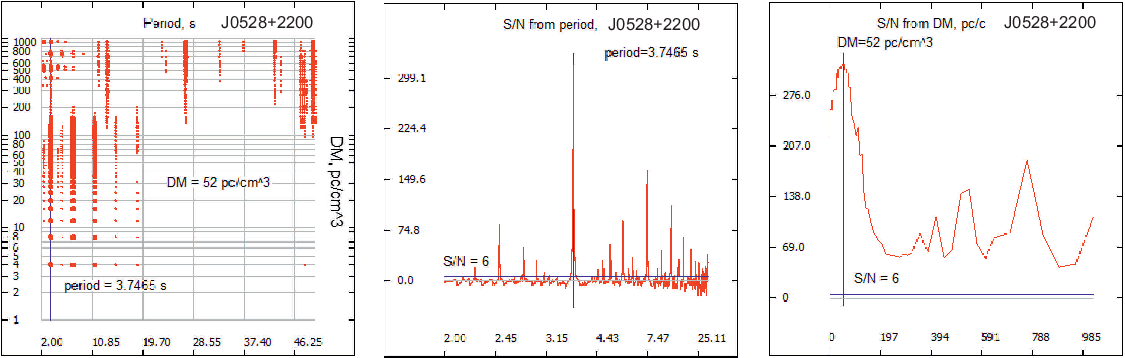}
	\caption{An example of a $P/DM$ map (left panel), a summed periodogram (middle panel), and the dependence of the harmonic maximum in the periodogram (its $S/N$) on $DM$ (right panel) generated by a data viewer for the famous pulsar J0528+2200. The intersection of the green and blue lines on the map indicates the $P$ and $DM$ of the strongest signal in the periodograms. The $P/DM$ map shows signals having $S/N > 50$.}
	\label{Fig1}
\end{figure*}

After the candidates are visually detected on the maps, they are searched through the ATNF catalog. If the source is not identified in the ATNF, work is carried out to show that a new source is being observed, or the visible harmonics are of a different nature. To do this, all candidates for new pulsars are tested in high-time-frequency resolution data.

The search algorithm is described in detail in \citeauthor{Tyulbashev2024} (\citeyear{Tyulbashev2024}). The bottom line is that the initial values of the candidate's $P$ and $DM$ are taken from the received maps. For each available day, hundreds of mean profiles are built with $P$ and $DM$ sorted next to the selected initial values. All processed days are sorted by decreasing the maximum $S/N$ in the estimated mean profile. This sort helps you choose the days with the best profiles. If the maxima in the average pulsar profiles for different observation sessions are obtained at close $P$ and $DM$, it is considered that a new pulsar has been detected. The values of $P$ and $DM$ at which the maximum $S/N$ is observed in the mean profile may differ from the initial values taken from the $P/DM$ maps. If the mean profile was obtained, these adjusted values are included in the tables.

The sensitivity of LPA3 in a standard session lasting 3.5 minutes makes it possible to detect pulsars with $S_{int} = 6-8$~mJy outside the Galactic plane and $S_{int} = 15-20$~mJy in the Galactic plane (\citeauthor{Tyulbashev2016}, \citeyear{Tyulbashev2016}). As noted above, with an improvement in sensitivity of $\sim 10$ times, pulsars with $S_{int} = 1.1-2.7$~mJy can be detected on $P/DM$ maps outside the plane and in the Galaxy plane. Therefore, if a weak pulsar does not have strong (5-10 times) fluctuations in the flux density on different days, during the verification search there is often not a single session for which the mean profile can be obtained (\citeauthor{Tyulbashev2024}, \citeyear{Tyulbashev2024}).

\section{Results}

A check in the ATNF catalog shows that the area has $+21^{\circ} < \delta < +42^{\circ}$ there are 14 canonical (having regular radiation) pulsars, in which $P > 2$~s, DM<200 pc/cm$^3$. 9 of the 14 pulsars (J0323+3944; J0349+2340; J0509+37; J0528+2200; J0546+2441; J0928+30; J0944+4106; J1746+2245; J1845+21) are found on the $P/DM$ maps. All these pulsars were previously detected and searched using summed Fourier power spectra. 3 of these 9 pulsars were previously discovered by us at LPA3\footnote{https://bsa-analytics.prao.ru/en/pulsars/new/}: J0509+37; J0928+30; J1845+21.

5 of the 14 pulsars (J1829+25, J2000+2920, J2011+3006, J2015+2524, J2111+4058) were not detected in the summed FFA spectra. Previously, the search in the same area using FFT spectra was carried out over an eight-year interval\footnote{https://bsa-analytics.prao.ru/en/pulsars/known /} (\citeauthor{Tyulbashev2024}, \citeyear{Tyulbashev2024}), and none of these 5 pulsars were detected either. At the same time, it is known that some pulsars exhibit flare behavior. These are intermittent pulsars, pulsars with nullings, pulsars with strong intrinsic variability and/or variability caused by the interstellar medium. The search for such pulsars in the summarized FFA spectra is not optimal, since the sessions in which the pulsar could be detected will be combined with sessions in which its radiation is absent or very weak. For such pulsars, it is necessary to search for radiation on individual observational sessions. For example, in \citeauthor{Tyulbashev2023} (\citeyear{Tyulbashev2023}), rotating radio transients (RRATs) J0642+30 and J1516+27 were noted, in which regular (periodic) radiation was not detected in the total power spectra with an accumulation equivalent to 6 days of continuous observations, but periodic radiation was detected in several separate sessions lasting 3.5 minutes.

We searched for five pulsars not found in the summarized periodograms in daily sessions over an interval of about 10 years (August 2014 - March 2024). None of the pulsars have been found. These pulsars were also not found in the search for pulsars at LOFAR (\citeauthor{Sanidas2019}, \citeyear{Sanidas2019}), where at a frequency of 135 MHz in one-hour sessions, the sensitivity reached 1.2 mJy. Estimation of the integral flux density of these undetected pulsars at a frequency of 111 MHz $S_{int} < 1$~mJy.

The area also includes 8 RRATs (J1336+33, J1354+2454, J1502+28, J1538+2345, J1940+2203, J1940+2231, J1946+24, J2008+3758). Of these, J1336+33, J1502+28 were previously discovered on LPA3\footnote{https://bsa-analytics.prao.ru/en/transients/rrat /}, and J1538+2345 was also detected by its pulses. Periodic emission of these RRATs was not detected in the total periodograms.

In addition to pulsars with periods of $P > 2$~s, 49 canonical (slow) pulsars in harmonics on multiple periods were found on the $P/DM$ maps. All of these pulsars have previously been detected and searched using summed power spectra. Therefore, in this paper we do not provide details for them, and note their discovery for the sake of completeness. As an example, we note the pulsar J2022+2854 (B2020+28), which has $P=0.3434$~s and is detected on a $P/DM$ map for a sevenfold period ($P = 2.4040$~s).

\begin{figure*}
	\includegraphics[width=\textwidth]{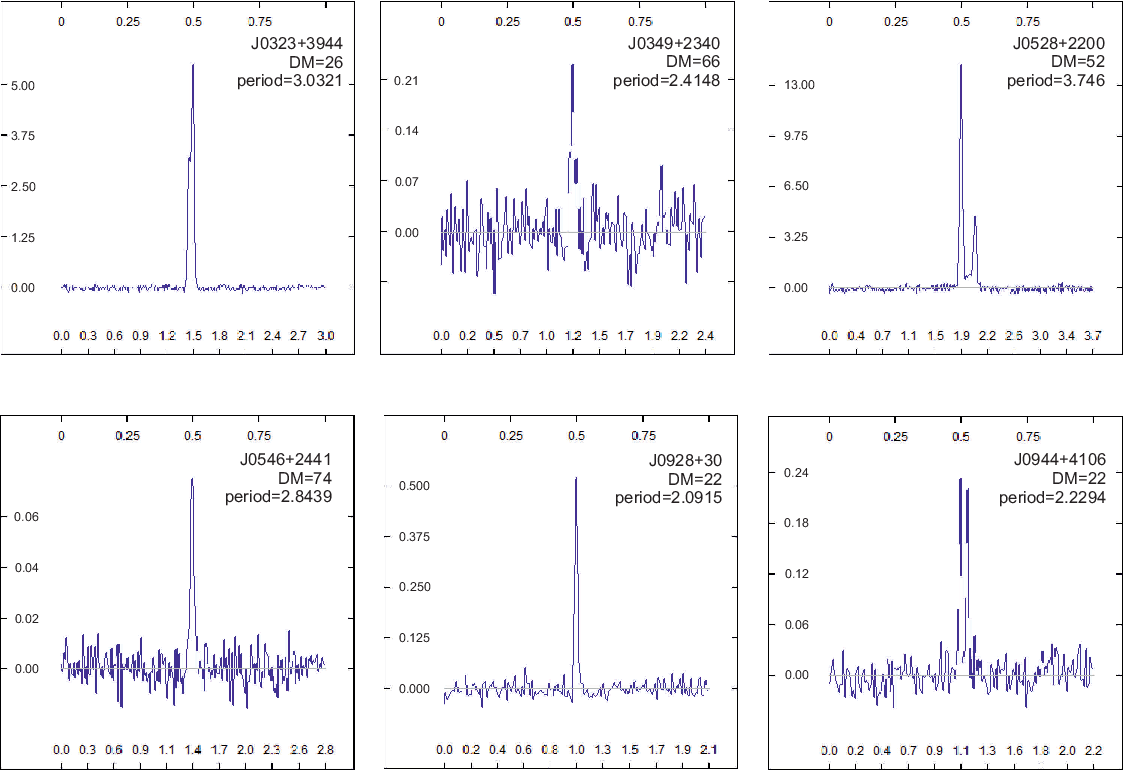}
	\caption{Mean pulsar profiles. The upper horizontal axis shows the phase, and the lower horizontal axis shows the time in seconds (the scale size is equal to the pulsar period). On the vertical axis, the flux density is in Jy. The maximum of the mean profile corresponds to phase 0.5.}
	\label{Fig2}
\end{figure*}

\begin{table*}
	\caption{Characteristics of pulsars trapped in the survey area}
	\begin{tabular}{|c|c|c|c|c|c|c|c|c|}
		\hline
		name & $P_{ATNF}, \, s$ & $DM_{ATNF}, \, pc/cm^3$ & $P_{LPA3}, \, s$ & $DM_{LPA3}, \, pc/cm^3$ & $S/N$ & $S_{peak}, \, Jy$ & $S_{int}, \, mJy$ & $W_{50}, \, ms$\\  		
		\hline
		J0323+3944 &  3.0320 & 26.1  &     3.0321 &  26    &    130  &  6012  &    115   &  65\\ 
		J0349+2340 &  2.4207 & 62.9  &     2.4204 &  65    &    8.5  &  296   &    4.3   &   29\\
		J0509+37*  &  2.4961 &  30   &      2.478 &  30    &      -  &   -    & $\le 1.6$&    -\\ 
		J0528+2200 &  3.7455 &  50.8 &     3.7448 &  50    &    82   & 14181  &   231    & 45\\
		J0546+2441 &  2.8438 &  73.8 &     2.8439 &  72    &     13  &   774  &    14.5  &  51\\
		J0928+30   &  2.0915 &  21.9 &     2.0915 &  22    &     33  &   576  &    7.7   &   29\\  
		J0944+4106 &  2.2294 &  21.4 &     2.2294 &  21    &     15  &   247  &    5.9   &    76\\
		J1746+2245* & 3.4650 &  49.8 &     3.4646 &  48    &      -  &      - &$\le 2.5$ &  -\\
		J1845+21*  &  3.7556 &   50  &     3.7203 &  40    &      -  &      - &$\le 1.6$ &  -\\
		\hline
	\end{tabular}
	\label{tab:1}
\end{table*}

Figure 2 shows the mean profiles of known pulsars, which were obtained from data with high time-frequency resolution. Table 1 shows the parameters of the pulsars. Columns 1-5 show the pulsar's name, $P$, and $DM$ according to the ATNF catalog and observations at LPA3. Columns 2-3 and 4-5 make it possible in practice to evaluate the accuracy of estimates based on observations on LPA3. Columns 6-9 show the $S/N$ in the profile, $S_{peak}$ (peak flux density), $S_{int}$, and profile widths at 0.5 of the profile height ($W_{50}$). When evaluating $S_{int}$ and $S_{peak}$, the mean profiles were selected, which visually had the best quality in the picture. Since the exact coordinates of the pulsars are in the ATNF, it is possible to make corrections related to the inaccurate hitting of pulsars in the center of the radiation pattern. Considering that the best quality recordings were used for the assessments, in the Table.1 contains estimates of the flux densities close to the maximum. An asterisk next to the pulsar's name indicates that the pulsar was detected in the total periodograms, but the mean profile from the individual observation sessions could not be obtained.

If the mean profile for the pulsar could not be obtained (pulsars J0509+37, J1746+2245, J1845+21), estimates based on $P/DM$ maps (based on the $S/N$ of the detected pulsar in the summarized periodogram) were placed in the table. The estimation of the integral flux density from the summed FFA spectra is based on the expected sensitivity of LPA3 in the direction of the pulsar, taking into account the temperature of the background of the Galaxy. The background temperature is recalculated from the temperature on the isophotes obtained at a frequency of 178 MHz (\citeauthor{Turtle1962}, \citeyear{Turtle1962}). The recalculation was performed under the assumption that the background temperature obeys the law of $T\sim \nu^{-2.55}$. If no pulsar was detected in the summarized periodograms, an upper estimate of the flux density was made based on the pulsar coordinate, background temperature, and expected sensitivity in the direction of the pulsar.

Estimates of $S_{int}$ pulsars from Table 1 obtained at close frequencies (103, 111, 120, 135, and 150 MHz) were previously published in \citeauthor{Malofeev2000} (\citeyear{Malofeev2000}); \citeauthor{Bilous2016} (\citeyear{Bilous2016}); \citeauthor{Tyulbashev2016} (\citeyear{Tyulbashev2016}); \citeauthor{Sanidas2019} (\citeyear{Sanidas2019}). For pulsars J0323+3944, J0528+2200, J0546+2441, J0944+4106, and J1746+2245, the estimates in Table 1 are consistent with those from earlier studies. For pulsars J0349+2340, J0509+37, J0928+30, and J1845+21, we were unable to find in the literature estimates of integral flux densities determined at frequencies of 100-150 MHz close to the frequency of observations at the LPA LPI. Estimates of peak and integral flux densities are given for the first time in the specified range.

In addition to the known pulsars, two new pulsars have been discovered (J1845+41, J1951+28). Figure 3 shows the drawings used for the visual detection of pulsars. Vertical segments are visible on the $P/DM$ maps, limiting the possible $DM$. Harmonics on multiple periods are also visible.

\begin{figure*}
	\includegraphics[width=\textwidth]{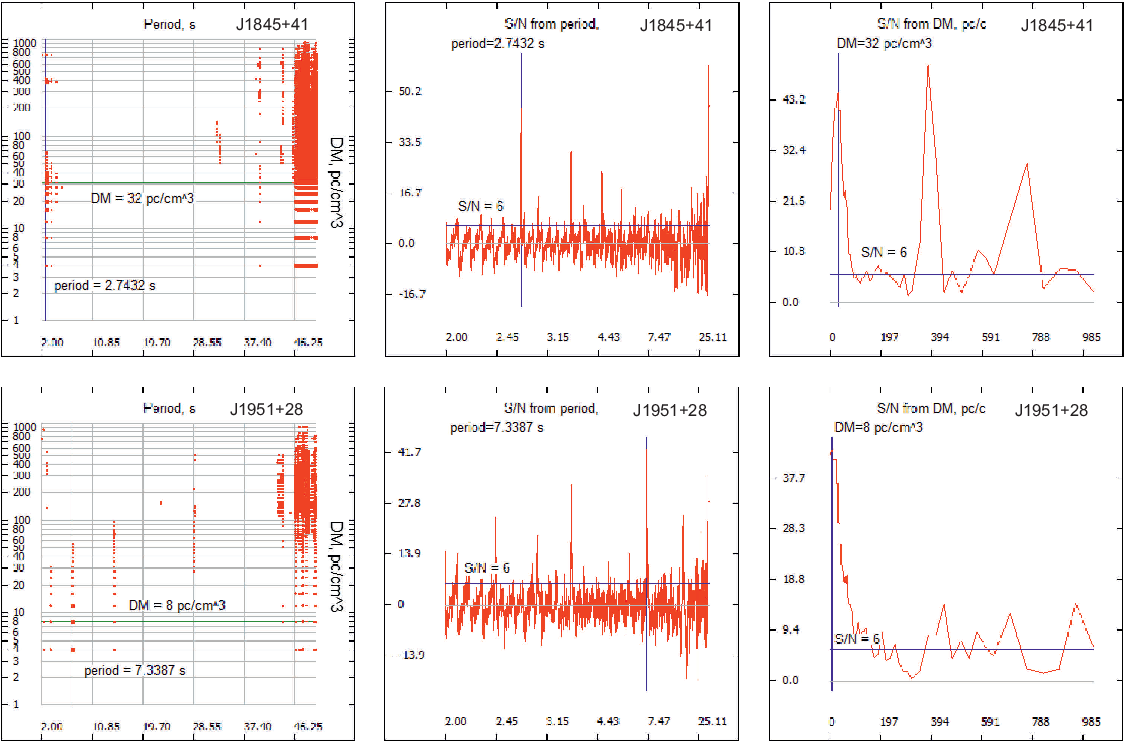}
	\caption{The panels repeat the panels in Fig.1 for the two new pulsars. The periodograms show multiple harmonics of pulsars. On the right side, the red stripes correspond to periods greater than 30s. No signals with large $S/N$ were detected in periodograms constructed for periods of 30-90s.}
	\label{Fig3}
\end{figure*}

The initial values of $P$ and $DM$ obtained by searching in periodograms are shown on the maps (Fig.3). For pulsar J1951+28, the pulsar parameters were refined from the mean profiles (see Fig.4) obtained during the enumeration of the initial values of $P$ and $DM$ taken from the $P/DM$ maps. The mean profile was obtained using data with high time-frequency resolution, as described in the section "Observations and processing".

\begin{figure}
	\centering
	\includegraphics[width=0.7\columnwidth]{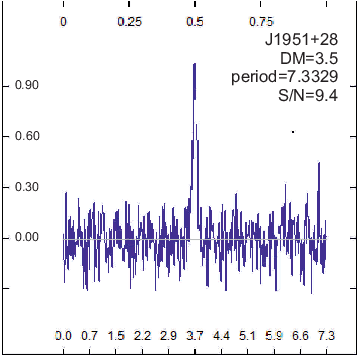}
	\caption{The mean profile of pulsar J1951+28. The axes are marked as in Fig.2.}
	\label{Fig4}
\end{figure}

Extracted characteristics of new pulsars:

\textbf{J1845+41}: $\alpha_{2000} = 18^h45^m40^s \pm 30^s$; $\delta_{2000} = 41^{\circ}13^\prime \pm 10^\prime$; $P = 2.7432 \pm 0.0050$~s; $DM = 30 \pm 3$~pc/cm$^3$; $S_{int} < 1$~mJy; $W_{50} < 100-200$~ms. 

\textbf{J1951+28}: $\alpha_{2000}=19^h51^m30^s \pm 45^s$; $\delta_{2000}=28^{\circ}46^\prime \pm 10^\prime$; $P=7.3342 \pm 0.0010$~s; $DM=3.5\pm 2$~ pc/cm$^3$;  $S_{peak}=0.68$~Jy; $S_{int}=11$~mJy; $W_{50}=130$~ms.

Pulsar J1951+28 is located close to the Galactic plane ($b<1^\circ$). Consequently, he repeatedly got into different surveys on the search for pulsars. The pulsar has high values of peak and integral flux density and, nevertheless, has not been detected so far. In our ten-year series of observations, he has dozens of sessions lasting 3.5 minutes, where we managed to obtain his mean profile. That is, the pulsar emits stably in the radio range. Note also that pulsar J1951+28 is one of the closest pulsars to the Sun with very low luminosity. The distance $D=152$~pc is determined for it using the YMW16 (\citeauthor{Yao2017}, \citeyear{Yao2017}) model. In the ATNF pulsar catalog, only 12 pulsars are located closer to the Sun than this pulsar. Its luminosity is $L_{111}=11\times 0.152^2=0.25$~mJy kpc$^2$. For the convenience of readers, we will transfer this luminosity to a frequency of 400 MHz, at which, out of 3,534 pulsars known in ATNF as of May 2024, there are luminosity estimates for 806 pulsars. Assuming a spectral index of 1.8, we obtain an integral flux density of $S_{400}=1.1$~mJy, and a luminosity of $L_{400}=1.1\times 0.152^2=0.025$~mJy kpc$^2$. To date, ATNF has recorded the lowest luminosities for pulsars J0307+7443, J0613+3731, and J1016-5345, which are 0.044, 0.055, and 0.047 mJy kpc$^2$, respectively. The luminosity rating of J1951+28 is 2 times lower. We also note that usually low radio luminosities (or the absence of detected radio emission) and long periods are characteristic of magnetars, but since we do not have a timing for pulsar J1951+28, there is no way to estimate the value of its magnetic field.

\section{Discussion and conclusion}

In the Introduction, we indicated long-period pulsars found in the radio range. They have periods ranging from tens of seconds to tens of minutes. Pulsars with extra-long periods are also found in ranges other than the radio range. For example, \citeauthor{Oliveira2020} (\citeyear{Oliveira2020}) talks about the detection of pulsar J2056-3014 with a period of 29.6~s in the X-ray and optical ranges, \citeauthor{Pelisoli2022} (\citeyear{Pelisoli2022}) - about the detection of pulsar J0240+1952 with a period of 24.9~s, in \citeauthor{Pelisoli2023} (\citeyear{Pelisoli2023}) - about the detection of pulsar J1912-4410 with a period of 319~s. The work \citeauthor{Lutovinov2022} (\citeyear{Lutovinov2022}) talks about the detection of the X-ray pulsar J2043+4438 with a period of 742~s. The values of the periods of pulsars detected in the X-ray range are comparable to the values of the periods of radio pulsars noted in the Introduction.

However, the nature of pulsars with extra-long periods may be different. According to the authors of the original papers, some of the sources found are neutron stars, and some are white dwarfs. At the same time, it should be noted that for objects with periods more than a few tens of seconds, finding out their nature is not an easy task.

Obviously, the minimum periods of white dwarfs are limited from below, since the centrifugal force should not exceed the gravitational force of matter on the surface of the star. The minimum known periods of white dwarfs are $P\simeq 25$~s (\citeauthor{Oliveira2020}, \citeyear{Oliveira2020}; \citeauthor{Pelisoli2022}, \citeyear{Pelisoli2022}) about $10^4$ times longer than the minimum periods of pulsars. In turn, the periods of radio pulsars should be limited from above, since when crossing the line of death (\citeauthor{Chen1993}, \citeyear{Chen1993}) they should switch from the state of radio pulsar to the state of neutron star without radio emission.

Thus, pulsars and white dwarf can mix in the sample of objects with extremely long periods (from tens of seconds) in the radio range. Some pulsars may have longer periods than those of white dwarf. However, the nature of the periodic radiation source (a neutron star or a white dwarf) is not clear for all detected objects. There is no answer to the question of whether white dwarfs can have coherent radio emission that can be detected on modern radio telescopes, and whether there are mechanisms that can ensure the radio emission of pulsars beyond the death line. The population synthesis carried out in \citeauthor{Rea2024} (\citeyear{Rea2024}) (see figures and Table 1 in the article) shows the possible existence of pulsars with long periods, but their number should be small. On the other hand, the same work shows that the number of white dwarf pulsars can be large, but the possible mechanism providing their radio emission is not clear.

Periodograms can be used to detect pulsars with long periods and obtain statistics on the number of such pulsars. Since the search for pulsars using periodograms is computationally intensive, the capabilities of modern computing technology are apparently not enough to re-process all available archival data with surveys on the search for pulsars.

Let's highlight two FFA attempts to search for pulsars in archived survey data conducted on the GMRT (\citeauthor{Singh2022}, \citeyear{Singh2022}; \citeauthor{Singh2023}, \citeyear{Singh2023}) aperture synthesis system and on the 64-meter radio telescope at Parks (\citeauthor{Wongphechauxsorn2024}, \citeyear{Wongphechauxsorn2024}). In \citeauthor{Singh2022} (\citeyear{Singh2022}); \citeauthor{Singh2023} (\citeyear{Singh2023}), an FFA search was performed for pulsars with periods up to 100~s (GHRSS survey) at a frequency of 327 MHz in a area measuring 2800 sq.deg. In \citeauthor{Wongphechauxsorn2024} (\citeyear{Wongphechauxsorn2024}), data from the Parks survey (HTRU survey) were taken and an FFA search for pulsars with periods up to 432~s was performed at a frequency of 1.35 GHz in a 1 sq.deg area (toward the Galaxy center). Of course, the area is 1 sq.deg. small. However, as the authors of \citeauthor{Wongphechauxsorn2024} (\citeyear{Wongphechauxsorn2024}) note, about 1,000 canonical pulsars can be observed in this direction, so the small viewing area is compensated by the large expected density of pulsars. The high scattering expected in the direction of the Galactic center when searching for pulsars with long periods is apparently not as dangerous as for canonical pulsars. This is due to the fact that pulsars with long periods have wide pulses, and the scattering itself at a frequency of 1.35 GHz is low compared to scattering at low frequencies. In other words, sensitivity losses for pulsars with long periods and wide mean profiles are not as critical as for ordinary pulsars with periods of a second or less.

The theoretical sensitivity of $S_{min}$ in the GMRT survey was 0.5 mJy for $S/N=5$  (\citeauthor{Bhattacharyya2016}, \citeyear{Bhattacharyya2016}). In the Parks survey, taking into account the direction to the Galactic center, $S_{min}=0.1$~mJy for $S/N=6$ (\citeauthor{Keith2010}, \citeyear{Keith2010}).

Let's compare these sensitivities with the sensitivity of the PUMPS survey conducted at LPA3. To do this, we recalculate the sensitivity to a frequency of 111 MHz, assuming a spectral index of $\alpha=1.8 (S\sim\nu^{-\alpha}$) (\citeauthor{Maron2000}, \citeyear{Maron2000}).

We get $S_{min-HTRU}=9$~mJy, $S_{min-GHRSS}=3.5$~mJy, $S_{min-PUMPS}= 0.5-1.5$~mJy. The last estimate is given for the PUMPS survey for high and low galactic latitudes, respectively, when averaging periodograms over a five-year interval, when pulsars hit exactly the beam of the LPA3. When the pulsar hits between the beams of LPA3, the sensitivity estimate of $S_{min-PUMPS}$ will deteriorate to 1-3 mJy.

As noted above, periodograms are good at identifying pulsars with narrow pulses and with long periods. If we take $P>2$~s over a long period (as in this paper), it turns out that no new pulsars have been found using periodograms in the search for GPRS and HTRU. Our search revealed only 2 new pulsars with $P>2$~s, and the values of their periods indicate that these are ordinary radio pulsars. The sensitivity achieved in the PUMPS survey is comparable to or higher than the sensitivity of early pulsar search surveys. Therefore, the absence of new pulsars for periods greater than 10~s shows that the total number of pulsars with long periods is small.

A contradictory picture is emerging. In the PUMPS survey, pulsars with periods of $10<P<90$~s and located at declinations of $+21^\circ<\delta<+42^\circ$ were not detected. That is, their total number in the sky is small. On the other hand, the probability of accidental detection of three objects in the radio range with extra-long periods of $P\sim 1$~min and $P\sim 20$~min (\citeauthor{Caleb2022}, \citeyear{Caleb2022}; \citeauthor{HurleyWalker2022}, \citeyear{HurleyWalker2022}; \citeauthor{HurleyWalker2023}, \citeyear{HurleyWalker2023}) is low. These super-long-period pulsar detections, in our opinion, indicate a large number of similar objects that should be detected by a specially organized search. Moreover, pulsars with periods of $P\sim 20$~min and peak pulse flux densities greater than 10 Jy have been detected at frequencies of 80 and 150 MHz. The frequency of observations at LPA3 is 111 MHz, which is in the middle of the 80-150 MHz range. Therefore, at any reasonable spectral index, such objects should be visible on LPA3. The only consistent explanation, in our opinion, is that the majority of pulsars with $P>10$~s have real periods of $P>90$~s, and therefore they were not detected in our search. In this case, our search took place in the valley of periods ($10<P<90$~s) when there were no radio pulsars, and white dwarfs with radio emission had not yet begun.

In conclusion, we note the main results of the search for long-period ($P>2$~s) pulsars on the LPA3 radio telescope in daily observations over an interval of 5 years. The search was performed using summed periodograms based on declinations of $+21^\circ<\delta<+42^\circ$. 9 out of 14 (64\% of those entering the area) pulsars known from the ATNF catalog were detected. Five undetected pulsars are known from observations in the decimeter range and have not been confirmed in the meter range. For these five pulsars, an additional search was conducted in separate sessions lasting 3.5 minutes. No pulsars have been detected again. It is possible that these are weak pulsars with flare activity, or pulsars with a cut-offs of the spectrum in the meter range. The area contains 8 known RRATs with periods of $P>2$~s. Their regular radiation has also not been detected. New pulsars J1845+41 and J1951+28 have been found. Pulsar J1951+28 turned out to be close to the Sun. It is also interesting for its record low luminosity. The absence of new pulsars for periods greater than 10~s indicates a small number of such pulsars in the sky.

\section*{Acknowledgements}
The research was carried out at the expense of the Russian Science Foundation (RSF) grant No. 22-12- 00236 (https://rscf.ru/project/22-12-00236 /). We also thank L.B. Potapova for her help in designing the paper.


\begin{thebibliography}{99}
\bibitem[Hewish et al.(1968)]{Hewish1968} Hewish, A., Bell, S.~J., Pilkington, J.~D.~H., et al.  Nature, \textbf{217}, 709 (1968). 
\bibitem[Lovelace \& Craft(1968)]{Lovelace1968} Lovelace, R.~V.~E. \& Craft, H.~D.  Nature, \textbf{220}, 875 (1968). 
\bibitem[Staelin(1969)]{Staelin1969} Staelin, D.~H.  IEEE Proceedings, \textbf{57}, 724 (1969). 
\bibitem[Morello et al.(2020)]{Morello2020} Morello, V., Barr, E.~D., Stappers, B.~W., et al.  MNRAS, \textbf{497}, 4654 (2020). 
\bibitem[Cameron et al.(2017)]{Cameron2017} Cameron, A.~D., Barr, E.~D., Champion, D.~J., et al.  MNRAS, \textbf{468}, 1994 (2017). 
\bibitem[Parent et al.(2018)]{Parent2018} Parent, E., Kaspi, V.~M., Ransom, S.~M., et al.  ApJ, \textbf{861}, 44 (2018). 
\bibitem[Singh et al.(2023)]{Singh2023} Singh, S., Roy, J., Bhattacharyya, B., et al.  ApJ, \textbf{944}, 54 (2023). 
\bibitem[Tyul'bashev et al.(2024)]{Tyulbashev2024} Tyul'bashev, S.~A., Tyul'basheva, G.~E., Kitaeva, M.~A., et al.  MNRAS, \textbf{528}, 2220 (2024). 
\bibitem[Manchester et al.(2005)]{Manchester2005} Manchester, R.~N., Hobbs, G.~B., Teoh, A., et al.  AJ, \textbf{129}, 1993 (2005). 
\bibitem[Caleb et al.(2022)]{Caleb2022} Caleb, M., Heywood, I., Rajwade, K., et al.  Nature Astronomy, \textbf{6}, 828 (2022). 
\bibitem[Tan et al.(2018)]{Tan2018} Tan, C. M., Bassa, C. G., Cooper, S., Dijkema, T. J., Esposito, P., Hessels, J. W. T., Kondratiev, V. I., Kramer, M., Michilli, D., Sanidas, S., Shimwell, T. W., Stappers, B. W., van Leeuwen, J., Cognard, I., Griessmeier, J.-M., Karastergiou, A., Keane, E. F., Sobey, C. \& Weltevrede, P.,  LOFAR Discovery of a 23.5s Radio Pulsar. ApJ, \textbf{866}, 54 (2018).
\bibitem[Hurley-Walker et al.(2022)]{HurleyWalker2022} Hurley-Walker, N., Zhang, X., Bahramian, A., et al. Nature, \textbf{601}, 526  (2022). 
\bibitem[Hurley-Walker et al.(2023)]{HurleyWalker2023} Hurley-Walker, N., Rea, N., McSweeney, S.~J., et al.  Nature, \textbf{619}, 487 (2023). 
\bibitem[Rea et al.(2024)]{Rea2024} Rea, N., Hurley-Walker, N., Pardo-Araujo, C., et al. ApJ, \textbf{961}, 214 (2024). 
\bibitem[Tyul'bashev et al.(2016)]{Tyulbashev2016} Tyul'bashev, S.~A., Tyul'bashev, V.~S., Oreshko, V.~V., et al.  Astronomy Reports, \textbf{60}, 220 (2016). 
\bibitem[Shishov et al.(2016)]{Shishov2016} Shishov, V.~I., Chashei, I.~V., Oreshko, V.~V., et al.  Astronomy Reports, \textbf{60}, 1067 (2016). 
\bibitem[Tyul'bashev et al.(2022)]{Tyulbashev2022} Tyul'bashev, S.~A., Kitaeva, M.~A., \& Tyul'basheva, G.~E.  MNRAS, \textbf{517}, 1112 (2022). 
\bibitem[Tyul'bashev et al.(2020)]{Tyulbashev2020} Tyul'bashev, S.~A., Kitaeva, M.~A., Tyul'bashev, V.~S., et al.  Astronomy Reports, \textbf{64}, 526 (2020). 
\bibitem[Tyul'bashev et al.(2023)]{Tyulbashev2023} Tyul'bashev, S.~A., Kitaeva, M.~A., Brylyakova, E.~A., et al.  Astronomy Letters, \textbf{49}, 533 (2023). 
\bibitem[Sanidas et al.(2019)]{Sanidas2019} Sanidas, S., Cooper, S., Bassa, C.~G., et al., A\&A, \textbf{626}, A104 (2019).
\bibitem[Turtle \& Baldwin (1962)]{Turtle1962} Turtle, A.~J. \& Baldwin, J.~E.  MNRAS, \textbf{124}, 459 (1962). 
\bibitem[Bilous et al.(2016)]{Bilous2016} Bilous, A.~V., Kondratiev, V.~I., Kramer, M., et al.  A\&A, \textbf{591}, A134 (2016).
\bibitem[Malofeev et al.(2000)]{Malofeev2000} Malofeev, V.~M., Malov, O.~I., \& Shchegoleva, N.~V., Astronomy Reports, \textbf{44}, 436 (2000).
\bibitem[Yao et al.(2017)]{Yao2017} Yao, J.~M., Manchester, R.~N., \& Wang, N., ApJ, \textbf{835}, 29 (2017).
\bibitem[Oliveira et al.(2020)]{Oliveira2020} Lopes de Oliveira, R., Bruch, A., Rodrigues, C.~V., et al.  ApJL, 898, L40 (2020). 
\bibitem[Pelisoli et al.(2022)]{Pelisoli2022} Pelisoli, I., Marsh, T.~R., Dhillon, V.~S., et al.  MNRAS, \textbf{509}, L31 (2022). 
\bibitem[Pelisoli et al.(2023)]{Pelisoli2023} Pelisoli, I., Marsh, T.~R., Buckley, D.~A.~H., et al.  Nature Astronomy, \textbf{7}, 931 (2023). 
\bibitem[Lutovinov et al.(2022)]{Lutovinov2022} Lutovinov, A.~A., Tsygankov, S.~S., Mereminskiy, I.~A., et al.  A\&A, \textbf{661}, A28 (2022). 
\bibitem[Chen \& Ruderman (1993)]{Chen1993} Chen, K. \& Ruderman, M.  ApJ, \textbf{402}, 264 (1993). 
\bibitem[Singh et al.(2022)]{Singh2022} Singh, S., Roy, J., Panda, U., et al.  ApJ, \textbf{934}, 138 (2022). 
\bibitem[Wongphechauxsorn et al.(2024)]{Wongphechauxsorn2024} Wongphechauxsorn, J., Champion, D.~J., Bailes, M., et al.  MNRAS, \textbf{527}, 3208 (2024). 
\bibitem[Bhattacharyya et al.(2016)]{Bhattacharyya2016} Bhattacharyya, B., Cooper, S., Malenta, M., et al.  ApJ, \textbf{817}, 130 (2016). 
\bibitem[Keith et al.(2010)]{Keith2010} Keith, M.~J., Jameson, A., van Straten, W., et al.  MNRAS, \textbf{409}, 619 (2010). 
\bibitem[Maron et al.(2000)]{Maron2000} Maron, O., Kijak, J., Kramer, M., et al.  A\&AS, \textbf{147}, 195, (2000). 
\end{thebibliography}
\end{document}